\documentclass{svproc}
\usepackage{url}

\usepackage{graphicx}
\begin{document}
\mainmatter              
%
\title{Collisional relaxation and dynamical scaling 
in multiparticle collisions dynamics}
\titlerunning{Relaxation and dynamical scaling in MPC dynamics}  
%
\author{S. Lepri\inst{1,5} \and 
H. Bufferand \inst{3} \and
G. Ciraolo \inst{3} \and \\ P. Di Cintio\inst{2,5} \and Ph. Ghendrih \inst{3} \and R. Livi\inst{4,5,1}}
\authorrunning{S. Lepri et al.} 
\tocauthor{S. Lepri, H. Bufferand, G. Ciraolo, P. Di Cintio, Ph. Ghendrih and R. Livi}
\institute{Consiglio Nazionale delle Ricerche, Istituto dei Sistemi Complessi \\
via Madonna del piano 10, I-50019 Sesto Fiorentino, Italy
\and
Consiglio Nazionale delle Ricerche, Istituto di Fisica Applicata ``Nello Carrara" \\
via Madonna del piano 10, I-50019 Sesto Fiorentino, Italy
\and
CEA, IRFM, F-13108 Saint-Paul-lez-Durance, France
\and
Dipartimento di Fisica e Astronomia and CSDC, Universit\'a di Firenze, \\
via G. Sansone 1, I-50019 Sesto Fiorentino, Italy
\and
Istituto Nazionale di Fisica Nucleare, Sezione di Firenze, \\
via G. Sansone 1, I-50019 Sesto Fiorentino, Italy
}
\maketitle             
\begin{abstract}
We present the Multi-Particle-Collision (MPC) dynamics approach to simulate properties
of low-dimensional systems. In particular, we illustrate the method for a simple
model: a one-dimensional gas of point particles interacting through stochastic collisions 
and admitting three conservation laws (density, momentum and energy). Motivated from problems in
fusion plasma physics, we consider an energy-dependent collision rate that accounts 
for the lower collisionality of high-energy particles. We study two 
problems: (i) the collisional relaxation to equilibrium starting from an off-equilibrium
state and (ii) the anomalous dynamical scaling of equilibrium time-dependent correlation
functions.  For problem (i), we demonstrate the existence of long-lived population
of suprathermal particles that propagate ballistically over a quasi-thermalized 
background. For (ii) we compare simulations with the predictions of nonlinear fluctuating hydrodynamics for the structure factors of density fluctuations.
Scaling analysis confirms the prediction that such model belong to 
the Kardar-Parisi-Zhang universality class. 
\end{abstract}
\section{Introduction}
Simulation of many-particle systems can be computationally very demanding, even for simple models.
This is challenging expecially when trying to measure asymptotic properties like the 
celebrated long-time tails of correlation functions in the thermodynamic limit \cite{Pomeau1975}.
Although molecular dynamics is the most natural choice, alternative approaches based 
on effective stochastic processes have been proposed both for computational efficiency
and also to get some insight in the general properties of non-equilibrium systems.
In this contribution we will briefly review the Multi-Particle-Collision (MPC) approach
which was originally proposed by Malevanets and Kapral \cite{Malevanets1998,1999JChPh.110.8605M,2004LNP...640..116M} in the context of mesoscopic dynamics of complex fluids (e.g. polymers in solution, colloidal fluids). In essence, it is based on a stochastic and {\it local} protocol that redistributes particle velocities, while preserving the global conserved quantities such as total energy, momentum and angular momentum.\\
\indent In this contribution, we will illustrate the method referring 
to the simple case of a one-dimensional fluid. Since we are interested to 
explore possible application of the method as a tool to investigate 
fusion plasma, we will introduce an energy-dependent collision rate that
mimics Coulombian interaction in a simple manner. We will consider two
problems: (i) the relaxation to equilibrium from a non-equilibrium initial 
state and (ii) the demonstration or
dynamical scaling of time-dependent correlation functions.\\
\indent Thermalization of many-particle system is a classic problem of non-equilibrium statistical
mechanics and kinetic theory. In the context of fusion plasma, the question is relevant
in the low-collisionality regime where non-equilibrium condition generate 
populations of suprathermal electrons and heavy tails in the velocity distribution function
\cite{1959PhRv..115..238D,1960PhRv..117..329D}. 
These fast particles modify heat and charge transport and thus the overall performance
of magnetic confinement devices \cite{2017NucFu..57k4002Z}.\\
\indent On the other hand, transport and dynamical scaling in low-dimensional models 
have been long investigated in the recent 
literature \cite{2003PhR...377....1L,DHARREV,Lepri2016}. The main findings is that    
many-particle systems with one or two spatial degrees of freedom show anomalous transport properties
signaled by the divergence of transport coefficients, like the thermal conductivity in the 
thermodynamic limit \cite{2003PhR...377....1L,DHARREV,Basile08,Lepri2016}. 
A way to detect this anomalous feature is to study dynamical scaling of equilibrium 
correlation functions and the corresponding dynamical scaling exponent $z$ (defined below) and 
seek for deviations from the usual diffusive behavior.
More recently, a complete description has been put forward within  
the Nonlinear Fluctuating Hydrodynamics (NFH) approach, proposed independently by van Beijeren 
\cite{2012PhRvL.108r0601V} and 
Spohn \cite{2014JSP...154.1191S,spohn2016fluctuating}. These authors have shown that the statistical properties of 1D nonlinear hydrodynamics with three conservation laws (e.g. total energy, momentum and number of particles) are essentially described by the fluctuating Burgers equation which can be mapped onto the well-known Kardar-Parisi-Zhang
(KPZ) equation for the stochastic growth of interfaces \cite{1986PhRvL..56..889K}. 
As a consequence, correlations of spontaneous fluctuations
are  characterized by the KPZ dynamical exponent $z=3/2$ in one-dimension.
The origin of the nontrivial dynamical exponents are to be traced back to the 
nonlinear interaction of long-wavelength modes. 
The results depends on the fact that the isolated system admits three conserved quantities
whose fluctuations are coupled. Models with a different number of conservation laws 
(e.g. two like in Ref.\cite{2015JSP...tmp...48S}) may belong to other universality
classes characterized by different dynamic exponents.     
A generalization to a arbitrary number of conserved
quantities has been discussed recently \cite{Popkov2015}. 
\section{Multi-Particle-Collision method}
\label{mpc}
The MPC simulation scheme (see Refs. \cite{2009acsa.book....1G,kapral08} for a detailed review) consists essentially in partitioning the system of $N_p$ particles in $N_c$ cells where the local center of mass 
 coordinates and velocity are computed and rotating particle velocities in the cell's center of mass frame are around a random axis. The rotation angles are assigned in a way that the invariant quantities are locally preserved (see e.g. \cite{2009acsa.book....1G,2018arXiv180101177C}). All particles are then propagated freely, or under the effect of an external force, if present.\\
\indent In the case of the one-dimensional fluid we are interested in, the above steps can be carried on as 
follows \cite{DiCintio2015}. Let us denote by $m_j$ and $v_j$ the mass and velocity of the $j$-th 
particle 
and by $N_i$ and the instantaneous number of particles 
inside each cell $i$ on which the system is coarse grained. The collision step amounts to assign 
random values to the velocities inside each cell, under the constraint of conserving, besides the particle number, the linear momentum $P_i$ and the kinetic energy $K_i$.  
In practice, we extract random samples $w_j$ from a Maxwellian distribution at the kinetic 
temperature of each cell, and let $v_{j,{\rm old}} \to v_{j,{\rm new}}=a_iw_j+b_i$, 
where $a_i$ and $b_i$ are the unknown cell-dependent coefficients determined by the conditions
\begin{eqnarray}\label{sist}
P_i&=&\sum_{j=1}^{N_i} m_jv_{j,{\rm old}}=\sum_{j=1}^{N_i} m_jv_{j,{\rm new}}
=\sum_{j=1}^{N_i} m_j(a_iw_j+b_i);\nonumber\\
K_i&=&\sum_{j=1}^{N_i} m_j\frac{v_{j,{\rm old}}^2}{2}=\sum_{j=1}^{N_i} m_j\frac{v_{j,{\rm new}}^2}{2}=
\sum_{j=1}^{N_i} m_j\frac{(a_iw_j+b_i)^2}{2},
\end{eqnarray}
Equations (\ref{sist}) constitute a linear system that can be solved 
for $a_i$ and $b_i$ analytically \cite{DiCintio2015,2018arXiv180101177C}.
Finally, the propagation step on the positions $r_j$ 
for a preassigned time interval $\Delta t$
is operated and the procedure repeats.\\
\indent The above collision procedure assumes implicitly that the velocity exchange is an instantaneous process that is not mediated by an effective potential. A further physical ingredient can be added assuming that 
the collision occurs at a given rate chosen to mimic some feature of the microscopic interaction.
An interesting example is encountered in the 
modelization of plasmas of charged particles were the rate can be fixed to capture the essence of the Coulombian scattering at low impact parameters (i.e. of the order of the cell size) \cite{Bufferand2010}. 
In the simulations presented here, we perform the above interaction step 
with a cell-dependent Coulomb-like interaction probability \cite{2013PhRvE..87b3102B,DiCintio2015,2018arXiv180101177C}
\begin{equation}\label{prob}
 \mathcal{P}_i=\frac{1}{1+\Gamma_i^{-2}},
\end{equation}
where $\Gamma_i$ is the plasma coupling parameter computed in cell $i$, relating the average Coulomb energy and the thermal energy $N_ik_BT_i=2K_i$, defined by
\begin{equation}
\Gamma_i=\frac{q^2}{4\pi\epsilon_0 ak_BT_i}.
\end{equation}
In the expression above $q$ is the particles charge, and $a$ a mean inter-particle distance realted to the inverse of average number density $\bar n$ and $\epsilon_0$ is the vacuum permittivity.\\
\indent Since the scope of this paper is to study and compare transport in low-dimensional models, we mostly limit ourselves to consider only one dimensional plasmas in a static neutralizing background with charge density $\rho_b$. In conditions where the neutrality is violated (e.g. when the number density $n$ is no longer uniform), the self-consistent electrostatic potential $\Phi$ can be included by
simultaneously solving the 1D Poisson equation
\begin{equation}\label{poisson}
\nabla^2\Phi(r)=-(qn(r)+\rho_b(r))/\epsilon_0
\end{equation}
by some standard finite-differences method. The resulting electric field is used to propagate 
the particles between each collision step.   
The dynamics can be further generalized to higher-dimensional charged fluids in a straightforward manner.
For instance, in Ref. \cite{DiCintio2017} a study of two-dimensional case has been considered in detail.
Moreover, the effects of the electromagnetic fields in higher dimensions can be implemented via particle-mesh schemes solving self-consistently the Maxwell equations on the grid.
\section{Relaxation to equilibrium}
In this section we present simulations of collisional relaxation from non-thermal initial states towards equilibrium. In a first set of numerical experiments we study the evolution of systems characterized by so-called waterbag initial conditions, whereby, positions and velocities $r$ and $v$ are initially distributed according to a phase-space distribution function of the form
\begin{equation}
f_0(r,v)=\mathcal{C}n\Theta(v_m-|v|).
\end{equation}\label{water}
In the expression above, $\Theta(x)$ is the Heaviside step function, $n$ is the particle number density constant over the periodic simulation domain $\left[0,L\right]$, and the normalization constant $\mathcal{C}$ is defined 
by the condition
\begin{equation}
\int_0^L\int_{-v_m}^{v_m}f_0(r,v){\rm d}v{\rm d}r=1.
\end{equation}
In all simulations presented hereafter the times are expressed in units of $t_{*}=2\pi/\Omega_P$, where $\Omega_P=\sqrt{q^2n/\epsilon_0m}$ is the plasma frequency of the system and we have used units such that $q=m=\epsilon_0=k_B=1$. To quantify the collisionality level within the fluid and compare different
simulation protocols, we define the 
global parameter $\Gamma$ as the average of the $\Gamma_i$ over all cells 
evaluated at $t=0$.
\\
\begin{figure}
\begin{center}
\includegraphics[width=0.95\textwidth]{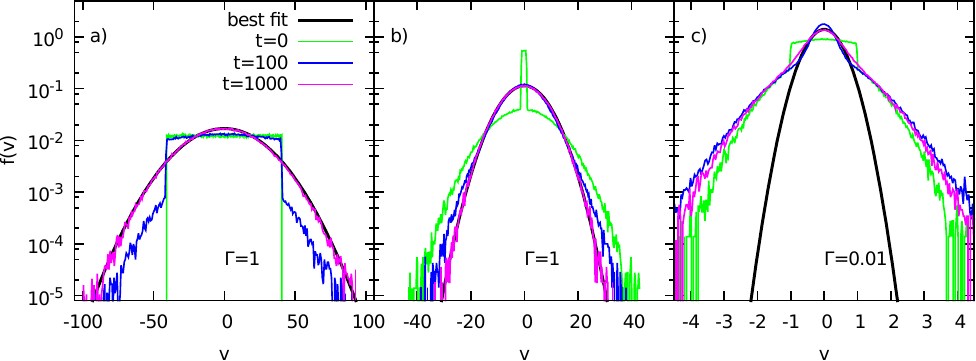}
\end{center}
\caption{Velocity distribution $f(v)$ at different times (thin lines) for a system of $N_p=2.5\times 10^5$ particles with waterbag initial conditions and $\Gamma=1$ (panel a), a system initially represented by the sum of a thermal and a waterbag distribution with an average $\Gamma=1$ (panel b), and a system initially represented by the sum of a thermal and a waterbag distribution with an average $\Gamma=0.01$ (panel c). In all cases the heavy solid line marks the best fit thermal distribution in the final state (in all cases $t=1000$).}
\label{relax}
\end{figure}
\begin{figure}
\begin{center}
\includegraphics[width=\columnwidth]{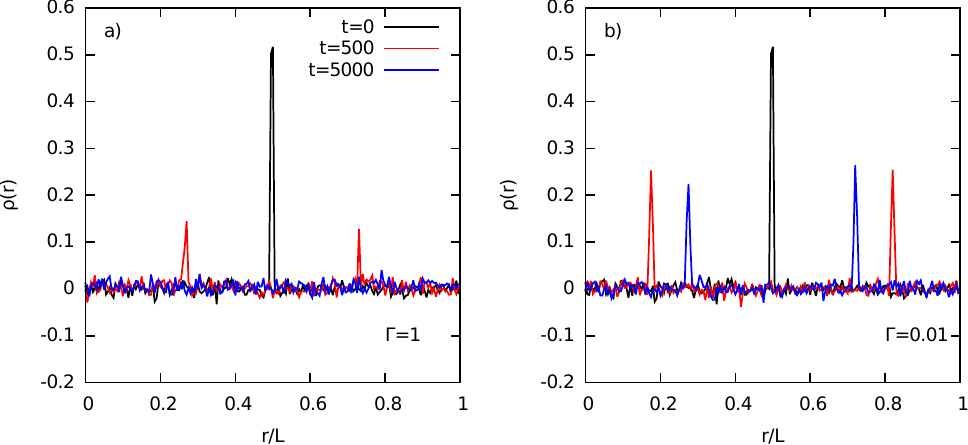}
\end{center}
\caption{Evolution of the charge density $\rho(r)$ for models characterized by an initially bunched supra-thermal population in a thermalized background with $\Gamma=1$ (panel a), and $\Gamma=0.01$ (panel b).}
\label{dens}
\end{figure}
\indent In Fig. \ref{relax}, panel a) we show the evolution of the velocity distribution $f(v)$ for a system with a combination of density and kinetic energy yielding an average coupling parameter $\Gamma=1$, and starting with a waterbag distribution. The (multi-particle) collisions gradually evolve $f(v)$ towards a Gaussian distribution (marked in figure by the heavy solid) while, due to the imposed neutrality, the tiny fluctuations of $\Phi$ play virtually no role. Remarkably, at intermediate times (here, $t=100$) $f(v)$ is characterized by high velocity thermal tails, while the bulk of the distribution for $-40<v<40$ still bears memory of the step-like initial $f_0(v)$. For larger values of $\Gamma$ (not shown here), $f(v)$ converges more and more rapidly to a thermal distribution. On the contrary, in the low-collisionality regime,  
for $\Gamma< 0.01$, the equilibrium is hardly reached on the simulation time. 
Indeed, particles in the high-velocity tails tend to decouple those belonging to the rest of the distribution
and perform an almost ballistic motion.\\
\indent In a second set of experiments we have studied the evolution of non-thermal populations in an already thermalized background system. In panels b) and c) of Fig. \ref{relax} we again show the evolution of $f(v)$ starting from an initial 
state constituted by two components one with $f_0$ given by Eq.(\ref{water}) and another with
\begin{equation}\label{thermal}
f_0(r,v)=\mathcal{C}n\exp(-v^2/2\sigma^2).
\end{equation}
While in the moderately coupled case (panel b, $\Gamma=1$) the two populations rapidly equilibrate, in the weakly coupled case (panel c, $\Gamma=0.1$) the final ($t=1000$) total velocity distribution features a low velocity region well fitted by a Gaussian distribution (heavy solid line) and high velocity power-law fat tails. 

Moreover, we have also  tested the stability of an initially localized bunch of mono-energetic particles in a thermalized background. In this set of numerical experiments, the initial conditions for the two populations were sampled from a thermal distribution like that of Eq. (\ref{thermal}), and a spatially bunched distribution of the form
\begin{equation}\label{bunch}
f_0(r,v)=\mathcal{C}\exp\left[-(r-r_*)^2/2s^2\right][\delta(v-v_*)+\delta(v+v_*)],
\end{equation}
where $\delta(x)$ is the Dirac delta function, $r_*$ is the centroid of the bunch, $s$ its width, and $v_*$ its velocity.\\
\indent In Fig. \ref{dens} we show the evolution of the charge density profile for an initially localized charge bunch placed in a periodic system of $10^6$ particles with $\Gamma=1$ (panel a) and 0.01 (panel b). In both cases, half of the $10^4$ bunch particles are initialized according to Eq. (\ref{bunch}) with $v=v_*=5\sigma$ and the other half with $v=-v_*$ in a gaussian bunch with $s=L/100$. As expected, in the less collisional cases ($\Gamma=0.01$, panel b), the bunch particles do not mix with the thermal background and (the two halves of) the bunch remain essentially coherent (at least for $t<10^4$, the simulation time), while for a more collisional system ($\Gamma=1$) the bunch is already completely dispersed at $t=5000$ by the interplay of collisions and mean field effects.
\section{Dynamical scaling}
As mentioned in the introduction, we are also interested in the scaling properties of time-dependent
correlation functions evaluated in some equilibrium ensemble (typically the 
microcanonical one). In the simulations aimed at computing equilibrium correlation functions, the 
initial conditions on position and velocity are extracted from Eq. (\ref{thermal}) and a uniform neutralizing background is assumed. For such distribution the local 
coupling parameters , 
Equation (\ref{prob}), are basically uniform over the whole system $\Gamma_i\approx \Gamma$. For this reason, and in order to save computational time, $\Gamma$ is evaluated at the beginning of the simulation and used as the single control parameter. Moreover, as in the limit of neutrality the electrostatic field vanishes, we do not solve Eq. (\ref{poisson}) and we simply impose $\nabla\Phi(r)=0$.\\
\indent The observables we will focus are the dynamical structure factors
of the conserved quantities defined at the resolution set by the cell partition.
Denoting  
$\xi_l$ as a shorthand notation for energy, momentum or density in the $l-$th cell 
($\mathcal{E},P,\rho$ respectively)
It is defined by first performing the discrete space-Fourier transform 
\begin{equation}
\hat \xi({k},t)=\frac{1}{N_c}\sum_{l=1}^{N_c} \xi_l\exp(-\imath {k}l).
\end{equation}
The dynamical structure factors $S_\xi(k,\omega)$ are defined as the modulus squared of the subsequent temporal Fourier transform  
\begin{equation}\label{somega}
S_\xi({k},\omega)=\langle|\hat \xi({k},\omega)|^2\rangle.
\end{equation}
Since we are working with periodic boundary conditions, the allowed values of 
the wave number $k$ are always integer multiples of $2\pi/N$, therefore in the rest of the paper 
we will sometimes refer to the (integer) normalized wave number $\tilde{k} = kN/2\pi$.
\\
\indent To connect with transport problems, we also considered the correlation function of 
the currents $J_\xi$, associated to the conserved quantity $\xi$. As above 
we choose to define the currents on the simulation grid
\begin{equation}
J_\xi(t)=\sum_{i=1}^{N_c}  \left[ \xi_i^{\prime}(t)-\xi^{\prime}_{i-1}(t-\Delta t)\right].
\end{equation}
Here, the prime is a shorthand notation to remind that only particles who moved from 
cell $i-1$ to $i$ between successive time steps must be considered in each term of the
sum.
We thus computed $C_\xi=\langle|\tilde J_\xi(\omega)|^2\rangle$ 
where the tilde denotes the Fourier transform in the time domain.\\
\indent According to the NFH theory \cite{2014JSP...154.1191S}, long-wavelength fluctuations are described in terms of hydrodynamic modes: in a system with three conserved quantities like chains of coupled oscillators with momentum conservation, the linear theory would yield two propagating sound modes and one diffusing 
heat mode, all of the three diffusively broadened. 
Nonlinear terms can be added and treated 
within the mode-coupling approximation \cite{Delfini07b,2014JSP...154.1191S}
that predicts that, at long times, the sound mode correlations 
satisfy Kardar-Parisi-Zhang scaling, while the heat mode correlations follow a L\'evy-walk scaling. As a consequence, it is expected that $S_\xi$ should be a 
combination of three modes correlations. For instance, for $k\to 0$, $S_\rho(k,\omega)$ should display sharp peaks at $\omega = \pm\omega_{\rm max}(k)$
that correspond to the propagation of sound modes 
and for $\omega\approx\pm \omega_{\rm max}$ it should behave as
\begin{equation}\label{scaling32}
S_\rho({k},\omega)\sim f_{\rm KPZ}\left(\frac{\omega\pm\omega_{\rm max}}{\lambda_s k^{3/2}}\right).
\end{equation}
Remarkably, the scaling function $f_{\rm KPZ}$ is universal and known exactly
 \cite{2014JSP...154.1191S} albeit not known in a closed form so that one
has to be evaluate it numerically \cite{private}. The nonuniversal coefficients 
$\lambda_s$ are model-dependent and, in principle,  can be evaluated in terms 
of static correlators \cite{2014JSP...154.1191S}.\\
\indent Another relevant signature of 
anomalous transport is the presence of long-time tails in the correlations or, 
equivalently, of a low frequency singularity. For instance, it is expected that 
$C_\mathcal{E}$ should diverge, in the large-size and low-frequency limits, 
as $\omega^{-1/3}$ \cite{2003PhR...377....1L,DHARREV,Lepri2016}.\\
\indent For a chain of coupled anharmonic oscillators with three conserved quantities like 
the Fermi-Pasta-Ulam chain, such theoretical predictions have been successfully 
compared with the numerics \cite{Das2014a,DiCintio2015}.
Other positive tests have been reported in Ref.\cite{Mendl2013}.
\begin{figure}
\begin{center}
\includegraphics[width=0.9\columnwidth]{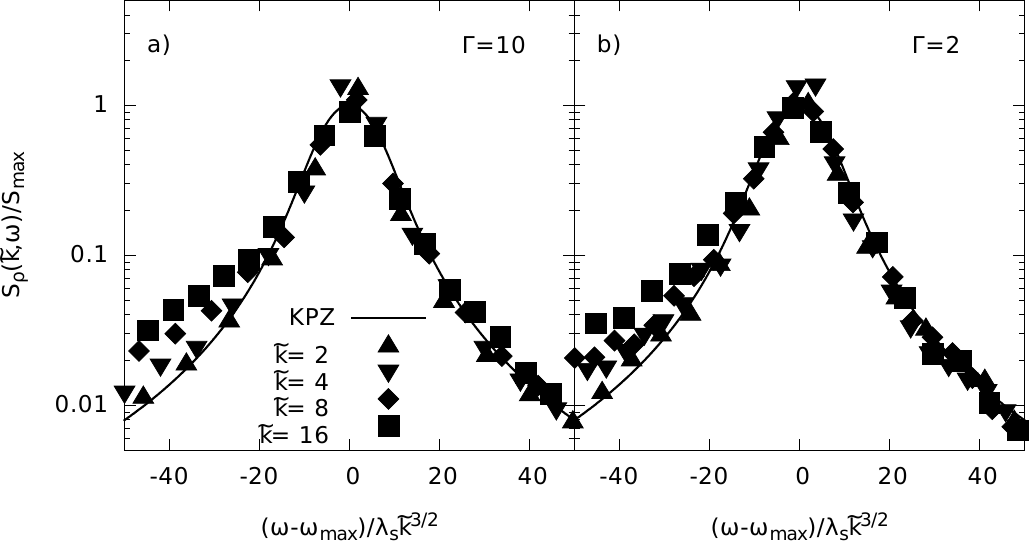}
\end{center}
\caption{Data collapse of the number density structure factors 
to the KPZ scaling function (solid line) of the Fourier spectra of the density profile modes with normalized wave number $\tilde{k}=2,$ 4, 8, and 16, for $\Gamma=10$ (panel a), and 2 (panel b). }
\label{scaling}
\end{figure}
\begin{figure}
\begin{center}
\includegraphics[width=0.8\textwidth]{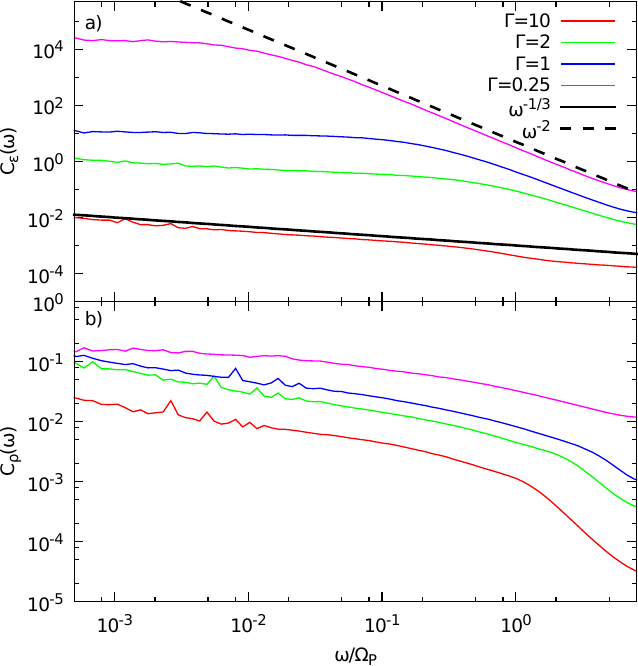}
\end{center}
\caption{For thermalized systems with $\Gamma=10,$ 2, 1 and 0.25: Fourier spectra $C_\mathcal{E}$ of the energy current (panel a) and Fourier spectra of the charge density current $C_{\rho}$ (panel b). The curves are averaged over 200 independent realizations. In all cases, the frequency $\omega$ is rescaled to the plasma frequency. The cross-over from the $\omega^{-1/3}$ to the $\omega^{-2}$ behavior of $C_\mathcal{E}$ at around $\Gamma=2$ is evident. To guide the eye, the dashed and solid black curves with the two slopes $-1/3$ and $-2$ have been added to the plot.}
\label{corrfunct}
\end{figure}
We have performed a series of numerical test for the MPC dynamics presented above.
Fig. \ref{scaling} shows the structure factors of density and energy for two strongly collisional cases with $\Gamma=10$ and 2, corresponding to relatively strong collisionality. 
Within statistical fluctuations the  data display a good data collapse and the lineshape fits  
with the KPZ-scaling function as predicted by Equation (\ref{scaling32}). 
It should be also mentioned that the same type of agreement has been shown to hold also 
for quasi-one-dimensional MPC dynamics, namely in the case of a fluid confined in a
box with a relatively large aspect ratio \cite{DiCintio2017}.
Another prediction of NFH is that the energy structure factors should display
a so-called L\'evy peak at zero frequency \cite{spohn2016fluctuating}. However, the data reported 
in \cite{DiCintio2015} (see in particular Fig. 6) show that the contribution of the sound
modes is pretty large, thus hindering the direct test of the prediction at least on the
timescales of such simulations. 
\\
\indent In Fig. \ref{corrfunct} we present the Fourier spectra $C_\mathcal{E}$ (panel a), and $C_\rho$ (panel b) of the energy and density currents, respectively, for four typical values of the ratio $\Gamma=10,$ 2, 1 and 0.25, and for $N_p=12000$ particles distributed on $N_c=1200$ cells. For strongly interacting systems (i.e. $\Gamma\geq 10$) one recovers the $\omega^{-1/3}$ behavior of the energy correlator $C_{\mathcal{E}}$. Increasing the particle specific kinetic $k_BT$ energy at fixed density $n$ (i.e. reducing $\Gamma$ and the collisionality of the system), $C_\mathcal{E}$ shows a more and more prominent flat region at low frequencies departing form the $\omega^{-1/3}$ trend, and a high frequency tail with slope $\omega^{-2}$.The cross-over from the $\omega^{-1/3}$ to the $\omega^{-2}$ behavior of $C_\mathcal{E}$ is evident at around $\Gamma=2$. A different behavior is instead found for the density correlator $C_{\rho}$, showing instead a $\omega^{-0.45}$ slope in the central part and a $\omega^{-2}$ tail at large $\omega$.\\
\indent The presence of the flat portion in $C_{\mathcal{E}}$ for $\omega\to 0$, could  
be naively interpreted as the restoration of normal conductivity.  
A similar regime where the decay of current correlations is faster (exponential) than the expected  
power-law decay has been reported for arrays of coupled oscillators \cite{2013PhRvE..87c2153C}
and it was argued that thermal conductivity could turn to a normal behavior in the low-energy 
regimes. Later studies \cite{Das2014} actually showed that this may be be rather due to 
strong finite-size effects. We thus argue that also our results, should be interpreted as 
such, although the physical origin of the effect is yet unexplained.
It is also puzzling that structure factors exhibit the scaling predicted by NFH over a wide range of values
of the control parameter $\Gamma$ whereby a clear crossover is seen in the current spectra upon reducing the collisionality of the particles (see again the panel a of Fig.\ref{corrfunct}). 
\section{Conclusions}
We have shown that the MPC method is a computationally convenient tool to study nonequilibrium
properties of many-particle systems. From the point of view of statistical mechanics, the 
models are relatively simple to allow for a detailed studies of basic problems like 
the ones discussed above. Despite its efficiency, the one-dimensional models is still 
affected by sizeable finite-size effects, particularly close to almost-integrable limits
of weak collisionality.\\ 
\indent Another attractive feature is that, introducing a suitable energy-dependent collision
probability allows to study, at least at a phenomenological level, some interesting 
issues of confined plasmas, like the effect of suprathermal particles. As a further 
development, interaction with external reservoirs exchanging energy and particles can
be included easily, thus allowing to study genuine nonequilibrium steady states.  
\section*{Acknowledgements}
This work has been carried out within the framework of the EUROfusion Consortium and has received funding from
the Euratom research and training program 2014-2018 under grant agreement No 633053 for the project WP17-ENR-
CEA-01 ESKAPE. The views and opinions expressed herein do not necessarily reflect those of the European Commission.
\bibliographystyle{spphys}
\bibliography{biblio.bib,heat}
\end{document}